\pdfoutput=1

\documentclass[11pt]{article}

\usepackage[final]{coling}

\usepackage{times}
\usepackage{latexsym}

\usepackage[T1]{fontenc}

\usepackage[utf8]{inputenc}

\usepackage{microtype}

\usepackage{inconsolata}

\usepackage{graphicx}
\usepackage{amsmath}
\usepackage{xcolor}
\usepackage{multirow}
\usepackage{makecell}

\title{Hawkes based Representation Learning for Reasoning over Scale-free Community-structured Temporal Knowledge Graphs}

 \author{Yuwei Du, Xinyue Liu\thanks{Corresponding author}, Wenxin Liang, Linlin Zong, Xianchao Zhang \\
         School of Software, Dalian University of Technology, China\\
         \{xyliu, wxliang, llzong, xczhang\}@dlut.edu.cn\\
         duyuwei2023@mail.dlut.edu.cn
         }

\begin{document}
\maketitle
\begin{abstract}
Temporal knowledge graph (TKG) reasoning has become a hot topic due to its great value in many practical tasks. The key to TKG reasoning is modeling the structural information and evolutional patterns of the TKGs.
While great efforts have been devoted to TKG reasoning, the structural and evolutional characteristics of real-world networks have not been considered. In the aspect of structure, real-world networks usually exhibit clear community structure and scale-free (long-tailed distribution) properties. In the aspect of evolution, the impact of an event decays with the time elapsing. In this paper, we propose a novel TKG reasoning model called Hawkes process-based Evolutional Representation Learning Network (HERLN), 
which learns structural information and evolutional patterns of a TKG simultaneously, considering the characteristics of real-world networks: community structure, scale-free and temporal decaying. 
First, we find communities in the input TKG to make the encoding get more similar intra-community embeddings.  Second, we design a Hawkes process-based relational graph convolutional network to cope with the event impact-decaying phenomenon. Third, we design a conditional decoding method to alleviate biases towards frequent entities caused by long-tailed distribution. 
Experimental results show that HERLN achieves significant improvements over the state-of-the-art models.
\end{abstract}

\section{Introduction}
Temporal Knowledge Graph (TKG) is a dynamic multi-relational graph used to record evolutionary events and knowledge in the real world. TKGs indicate facts as quadruples (subject, relation, object, time) and are actually sequences of temporal subgraphs divided by the time (timestamp) dimension. Reasoning over TKGs aims to infer the missing quadruple facts, which has two settings: interpolation and extrapolation. Given a TKG with timestamps from $t_0$ to $t_K$, interpolation aims at inferring missing facts that occur at time $t$, where $t_0 < t < t_K$. Oppositely, extrapolation attempts to predict facts that occur at time $t$ with $t > t_K$.  In this paper, we focus on the extrapolation setting, which has gained much attention in recent years due to its great practical value in event prediction, question answering, and other areas \citep{KGAT, QuestionAnswer}. There are mainly two extrapolation tasks: entity prediction and relation prediction. We aim to propose a unified model that can accomplish both the entity and relation prediction tasks. 

The key to TKG reasoning is modeling the structural information and evolutional patterns of the TKGs. The prior extrapolation TKG reasoning models such as CyGNet \citep{CyGNet} and CENET \citep{CENET}, learn the evolutional patterns by generating the historical event vocabulary to predict repetitive events. Later models such as RE-GCN \citep{REGCN}, HisMatch \citep{HiSMatch} and HGLS \citep{HGLS}, employ a relational graph convolutional network (RGCN) \citep{RGCN} to capture the structural information from historical snapshots and use a recurrent neural network (RNN) to model the evolutional patterns. Some recent works such as TITer \citep{TITer} and DREAM \citep{DREAM} introduce reinforcement learning on the TKG reasoning task. 

Nevertheless, while TKGs are reflections of the real world, the structural and evolutional characteristics of real-world networks have not been considered in previous models. In the aspect of structure, real-world networks (e.g., social-networks) usually exhibit clear community structure and scale-free (long-tailed distribution) properties \citep{Networks}. In the aspect of evolution, the impact of an event decays with the time elapsing \citep{Hawkes}. Taking these characteristics into consideration not only can improve the reasoning performance, but also better facilitate the down-stream tasks. 

In this paper, we propose a novel TKG reasoning model called Hawkes process-based Evolutional Representation Learning Network (HERLN), which learns structural information and evolutional patterns of a TKG simultaneously, considering the characteristics from real-world networks: community structure, scale-free and temporal decaying. Specifically, our model consists of three modules: an embedding initializing module, an evolution encoding module and a conditional decoding module.

In the embedding initializing module, to exploit the community structure properties of TKGs, we first find communities in the input TKG, and apply a graph convolution network to get embeddings of events within each community. The embeddings are then used as inputs in the evolution encoding module, which make the evolution encoding module output more similar intra-community embeddings.
In the evolution encoding module, to cope with the event impact-decaying phenomenon, we design a Hawkes process-based relational graph convolutional network (HRGCN). The graph convolutional network contracts the structural information of the TKG, while the Hawkes process assigns different weights to the timestamps such that the impacts of events decay over time.
In the conditional decoding module, to alleviate biases towards frequent entities caused by long-tailed distribution, we construct a conditional decoder which consists of a hyper network and a query-specific decoder. The hyper network adjusts the parameters according to the query events and the decoder generates conditional intensity scores for the candidate entities based on the adjusted parameters.

The main contributions of this work are summarized as follows: 

\begin{itemize}
\item We recognize that the TKGs possess the structural and evolutional characteristics inherited from real-world networks: community structure, scale-free and temporal decaying, but they have not been considered or well exploited in previous studies.

\item We propose a Hawkes process-based evolutional representation learning network (HERLN), which consists of three modules: (1) An embedding initialize module, which extracts semantic information of community structure; (2) An evolution encoding module, which addresses the temporal decaying of event impact; (3) A conditional decode module, which alleviates the biases towards frequent entities caused by long-tailed distribution.

\item Our proposed model HERLN can predict entities and relations at the same time. Experimental results on four benchmark TKG datasets show that HERLN achieves significant improvements over the state-of-the-art models.
\end{itemize}

\section{Related work}

\subsection{Embedding-based methods}
Embedding-based methods encode the whole or part of the TKGs to obtain the embeddings of entities and relations, and use the embeddings to evaluate the possibility of missing facts. 

RE-NET \citep{RENET} proposes an auto-regressive architecture which uses a graph neural network (GNN) to capture local entity embeddings and a RNN to model interactions between entities over time. RE-GCN \citep{REGCN} constructs a static graph to get the static attributes and presents a framework that can execute both entity and relation reasoning. 

EvoKG \citep{EvoKG} uses an auto-regressive architecture and captures the ever-changing structural and temporal dynamics via recurrent event modeling. HiSMatch \citep{HiSMatch} generates background graphs, entity-related graphs and relation-related graphs to jointly model the evolutional patterns. HGLS \citep{HGLS} models local snapshots or global graphs by using different GNNs and decodes them to get the predicting scores. CENET \citep{CENET} combines historical and non-historical information and identifies highly related entities via contrastive learning. TARGAT \citep{TARGAT} captures the interactions of multi-facts at different timestamps. DLGR \citep{DLGR} learns the local and global perspective representations in a contrastive manner. DSTKG \citep{DSTKG} introduces two latent variables to capture the dynamic and static characteristics of entities in TKGs. 

$L^2$TKG \citep{L2TKG} finds the missing relationships on the known KGs first and then reasons on the completed graph and the original graph jointly. RETIA \citep{RETIA} constructs a hyper-graph to connect different relations in a high-dimensional space. 
These models use RNNs to represent the temporal information. Thus they are based on an assumption that the temporal sequences are equidistant, which is inconsistent with many real-life event sequences \citep{GHT}.

\subsection{Path-based methods}
Path-based methods find several related paths of query facts and select the most relevant one as the answer. 
TITer \citep{TITer} adopts reinforcement learning to sample actions from query-related trajectories based on a time-shaped reward function. xERTE \citep{xERTE} samples and prunes the query-related subgraph according to query-dependent attention scores.  TANGO \citep{NeuralODE} explores the neural ordinary differential equation to build a continuous-time model. TLogic \citep{TLogic} automatically mines recurrent temporal logic rules by extracting temporal random walks. 
DREAM \citep{DREAM} use generative adversarial networks to design an adaptive reward function.
However, the path-based methods focus on the local structure graph of the query, ignore the potential connection of events, and do not perform well on long-term reasoning.

In addition, it is worth noting that methods such as xERTE \citep{xERTE} and HISMatch\citep{HiSMatch} consider the impact of temporal information on prediction results. They encode timestamps and concat timestamps with entity embeddings. However, different from our work, they actually learns that entities with different time intervals have different impacts on results, rather than considering the gradual decay of event impacts.

\subsection{Hawkes process-based methods}
The Hawkes process \citep{Hawkes} is a stochastic process that models sequential discrete events occurring in continuous time. There are several works that combine the Hawkes process and neural networks for TKG reasoning. Know-Evolve \citep{KE} introduces a temporal point process to model facts evolved in the continuous time domain. GHNN \citep{GHNN} proposes a graph Hawkes process to capture the potential temporal dependence across different timestamps. However, Know-Evolve and GHNN do not use the graph structural information. GHT \citep{GHT} uses a temporal Transformer to capture long-term and short-term information jointly. However, none of the previous work has considered  problem of temporal decaying of events' impacts. Our proposed module uses the Hawkes process to assign different weights to the timestamps during message passing, thus the information of the event impact-decaying is encoded and utilized.

\section{Problem Formulation}
A temporal knowledge graph (TKG) $G=\{\mathcal{E},\ \mathcal{R},\ \mathcal{T},\ \mathcal{F}\}$ is a directed multi-relational graph, where $\mathcal{E},\ \mathcal{R},\ \mathcal{T} and \ \mathcal{F}$ denote the sets of entities, relations, timestamps and facts, respectively. 
A node in $G$ represents an entity $i \in \mathcal{E}$, and an edge $e_{ij}$ represents the interaction between node $i$ and node $j$ with relation $r \in \mathcal{R}$ at timestamp $t\in \mathcal{T}$. A fact in $G$ is a quadruple $q=(s, r, o, t)$ that represents a real-world event consisting of the relation $r$ between a subject entity $s$ and an object entity $o$ at timestamp $t$. 

Given a TKG $G_{[t_1: t_k]} = \{\mathcal{E},\ \mathcal{R},\ \mathcal{T},\ \mathcal{F}\  |\  \mathcal{T} = [t_1, t_k] \}$, the extrapolation reasoning task is to predict object $o_q$ in a query like $(s_q, r_q, ?, t_q)$ where $t_q > t_k$, or predict relation $r_q$ in a query like $(s_q, ?, o_q, t_q)$ where $t_q > t_k$.



\section{Method}

\begin{figure*}[thbp]
\centering
\includegraphics[width=2.0\columnwidth]{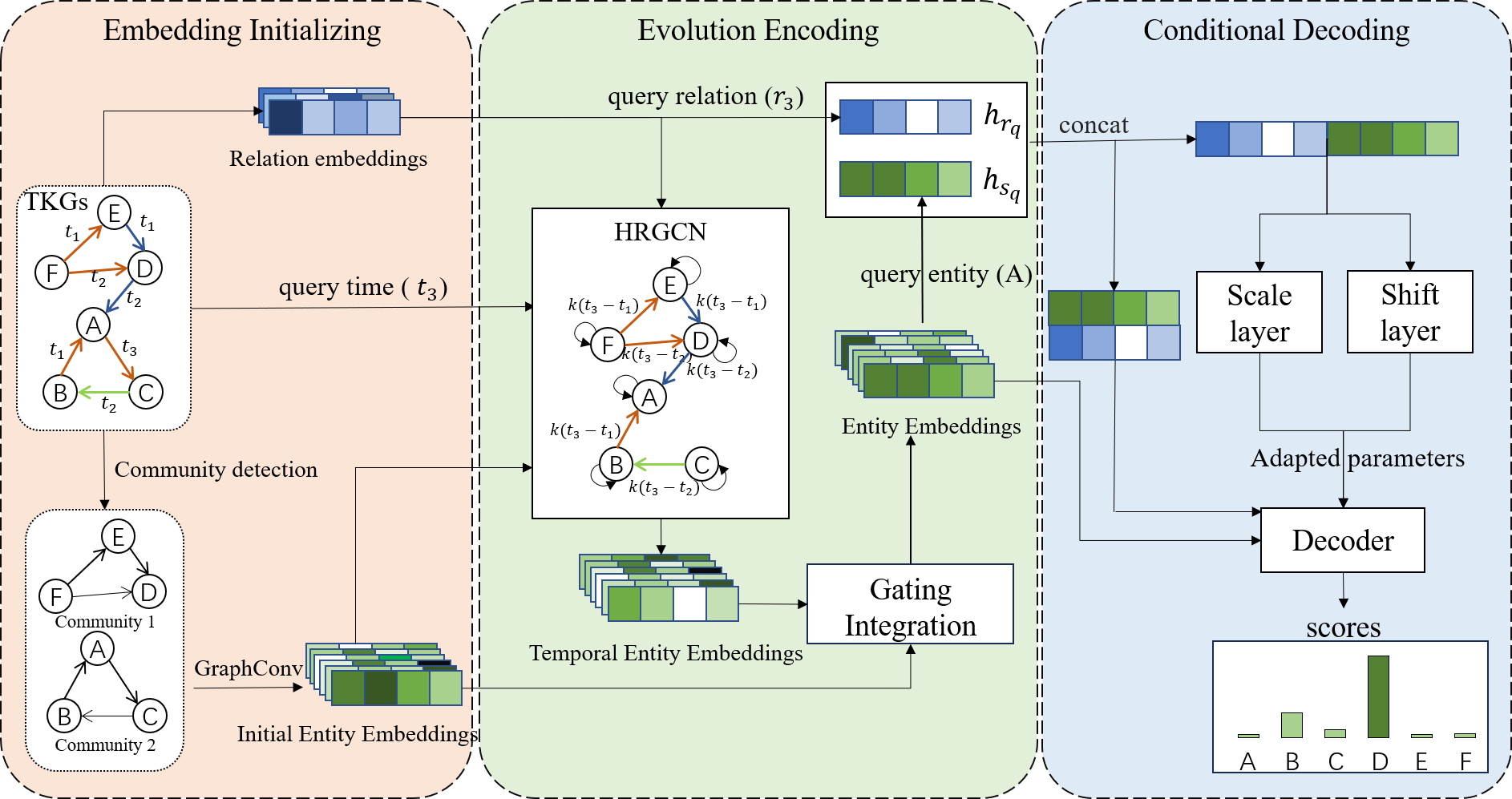}
\caption{Overall framework of the proposed HERLN model. HERLN is consists of three modules: an embedding initializing module, an evolution encoding module and a conditional decoding module. First, the embedding initializing module extracts the community structural information in input TKGs to get the initialized embeddings. Then the evolution encoding module updates the embeddings with the candidate-related historical structure to learn the evolutional patterns of events. Finally, the conditional decoding module reasons according to the embeddings and gets scores of the candidate entities, then select the entity with the highest score as the results. The input TKG contains 6 nodes (marked as A to F), three different types of edges (indicated in red, blue and green respectively) and three timestamps (from $t_1$ to $t_3$). Additionally, the model receives an incomplete quadruple as query. The query given in the figure is $(A,r_3,?,t_3)$. }
\label{fig}
\end{figure*}

The proposed Hawkes process-based evolutional representation learning network (HERLN) model is shown in Fig. \ref{fig}, it consists of three modules, an embedding initializing module, an evolution encoding module and a conditional decoding module. 
The embedding initializing module detects communities in the input TKG and embeds the interaction frequencies between entities within each community into the initialized entity embeddings.
With the initialized entity embeddings as input, the evolution encoding module updates the entity embeddings by learning the structural information from historical graph. 
The conditional decoding module uses representations of query quadruple to adjust the parameters and generate scores for candidates. 

\subsection{Embedding Initializing Module}

To initialize the embeddings of entities, we first identify the communities, and then extract semantic information in the communities to get the initialized entity embedding matrix $\mathrm{H}_C$. 

\subsubsection{Community Detection}
The interactions between entities in real world exhibit distinct community structures. Exploiting of the community structural properties contributes to the improvement of reasoning performance. For example, an entity that cooperates with a country like America is more likely to be a government or an organization rather than the citizens of a country. This information could be used to reduce the scores of entities that are not consistent with the facts.

We use a community detection algorithm on the entire TKG, which divides the entities set into different communities according to interaction frequencies, and obtains a graph that only contains inner-community links. The algorithm assigns each entity $i$ to its community $c_i$, and there are a total of $K$ communities in the TKG. 

In TKGs, there are no inherent community labels. Therefore, we require an unsupervised and reliable method to detect possible communities in the TKGs. And since a TKG is a multi-relational graph, we extend the Louvain algorithm \citep{Louvain} to handle with the multi-relational links.
Specifically, we calculate modularity $Q_r$ for different relation $r$ by Eq. \ref{Q_value}, which is an indicator that measures the quality of community detection.

\begin{equation}
\begin{aligned}
    Q_r &=\sum_{c}{[\frac{\sum in}{2m} - (\frac{\sum tot}{2m})^2]} 
    &= \sum_{c}{[e_c-a^{2}_{c} ]}   \label{Q_value}
\end{aligned}
\end{equation}
where $\sum in$ is the sum of weights of inner-community edges with relation $r$; $\sum tot$ is the sum of weights of all the edges with relation $r$ and $m$ is the total weight of edges on the whole graph. 

During the optimization process, when a community is merged into another community, the algorithm will calculate the modularity of the new entire graph, compare it with the modularity before merging to get $\Delta Q$ as described in Eq. \ref{Delta_Q}. 
\begin{equation}
\begin{aligned}
    \Delta Q &= \sum_{r} {\Delta Q_r}  \\  
    &=\sum_{r} \Big([\frac{\sum in +k_{i,in}}{2m}-(\frac{\sum tot + k_i}{2m})^2]     \\
    & \ \ - [\frac{\sum in}{2m} - (\frac{\sum tot}{2m})^2 - (\frac{k_i}{2m})^2]\Big)   \\
    &= \frac{1}{2m}(k_{i,in} - \frac{\sum_{tot}k_{i}}{m})  \label{Delta_Q}
\end{aligned}
\end{equation}
where $k_{i,in}$ is the sum of the edges' weights between node $i$ and the new community $in$; $k_{i}$ is the sum of the edges' weights between node $i$ and all the nodes in the graph. The algorithm ends when $\Delta Q$ no longer changes.
 
\subsubsection{Embedding Initialization}

In order to import the information contained in the communities into the embeddings, we use a GCN to generate embeddings on the community subgraphs, which is formalized as:
\begin{equation}
    h_i = \sigma\Big(\sum\limits_{j\in\mathcal{N}_i}\frac{1}{|\mathcal{N}_i|}Wh_j^{init} \delta(c_i,c_j) + W_0 h_i^{init}\Big)
    \label{gcn}
\end{equation}
where $h_i^{init}$ and $h_j^{init}$ are randomly initialized embeddings of nodes $i,j$; $W$ is the parameter of message passing between nodes; $W_0$ is the parameter of self updating of a node; $\delta(c_i,c_j)$ is an indicator which is set to 1 if $i$ and $j$ belong to a same community and 0 otherwise; $\sigma()$ is an activate function; $\mathcal{N}_i$ is the set of neighbors of node $i$.

\subsection{Evolution Encoding}
After getting the initialized embedding matrix, the next move is to encode the candidate-related historical structure to learn the evolutional patterns of events. In real-world events, the entity's status changes over time. Another common real-world phenomenon is that the impact of a event decays over time,
The entity embeddings should be able to cope with the variations. 
This module achieves these points by updating the entity embeddings with historical information.

\begin{figure*}[thbp]
\centering
\includegraphics[width=1.8\columnwidth]{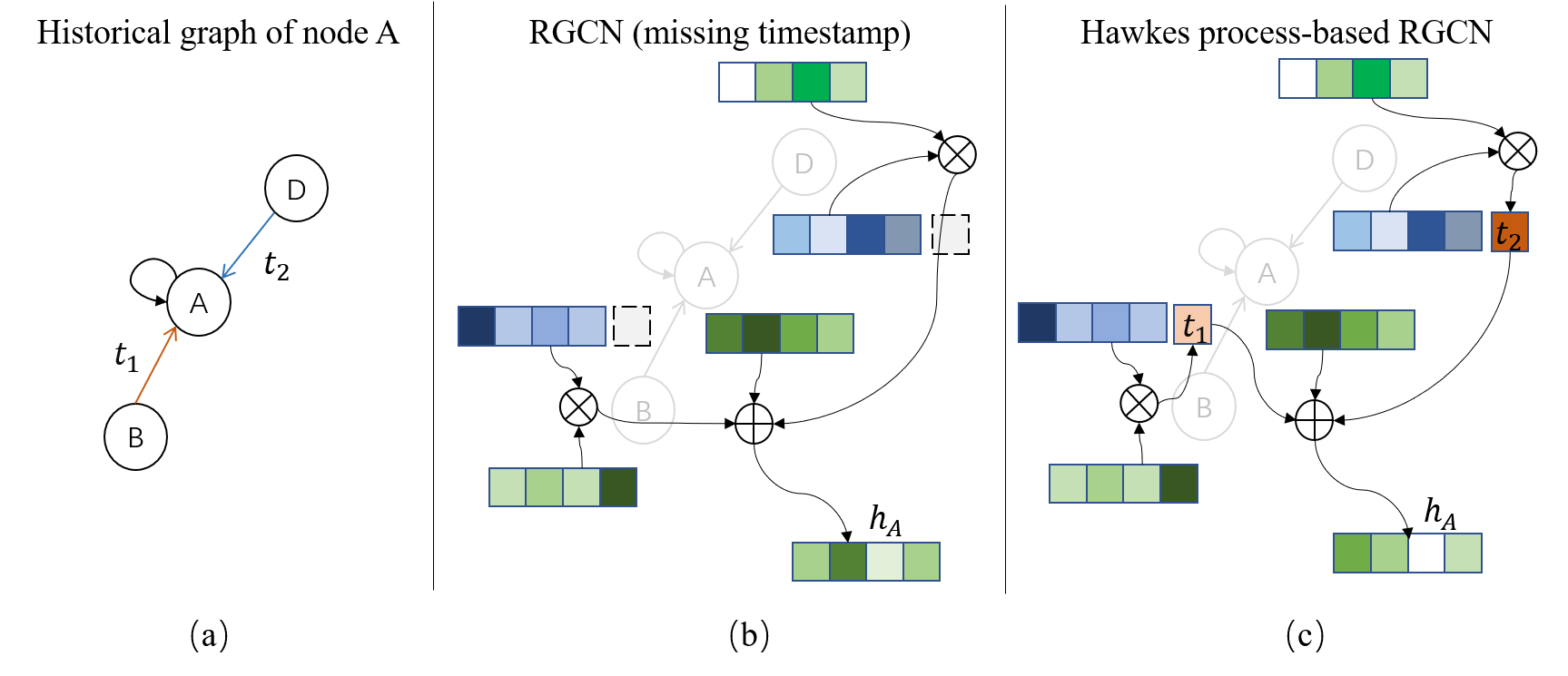}
\caption{The update process of HRGCN. (a) is a historical structure of node A, (b) shows the update process of a traditional RGCN, which does not use timestamp information and (c) shows the update process of HRGCN, which takes the time interval as decaying weight to represent the event declines over time.}
\label{FigHRGCN}
\end{figure*}

\subsubsection{The Hawkes Process on TKGs}
\label{sec:The Hawkes Process on TKGs}
The Hawkes process is a stochastic process that models a sequential discrete events that occur chronologically, which is typically modeled by a conditional intensity function. The intensity function $\lambda(t)$ represents the probability that events happen at $t$, it is defined as follows:

\begin{equation}
\begin{aligned}
    \lambda_{(s,r,o)}(t) =& \sum_{(s',r',o,t')\in \mathcal{F}_o }\gamma_{s', r'}(t')\kappa(t-t') \\
    &+ \mu_{s,r,o}(t) \label{hawkes}
\end{aligned}
\end{equation}
where $\mu_{s,r,o}(t)$ is the base intensity at time $t$;  $\mathcal{F}_o$ is the set of historical events related to node $o$; $\gamma(\dot)$ represents the amount of excitement induced by the corresponding events on results and $\kappa()$ is a kernel function to model the effect of historical events on the current event. 

To integrate the Hawkes Process into the TKG, we treat the Hawkes Process as an encoder-decoder network. For encoding, we use a function $En()$ to get representations $h_s, h_r$ and $h_o$ derived from their historical neighbors’ information, as shown in Eq. \ref{encoder}. For decoding, we transfer the representations into a certain intensity value with an appropriate decoding function $De()$, as shown in Eq. \ref{decoder}.
\begin{equation}
    h_s, h_r, h_o = En(G) \label{encoder}
\end{equation}
\begin{equation}
    \lambda_{(s,r,o)}(t) = De(h_s, h_r, h_o) \label{decoder}
\end{equation}

We present detailed implementations of $En()$ and $De()$ in the following sections.

\subsubsection{Encoding with Hawkes process-based RGCN}
To get the historical representation of entities described in Eq.\ref{encoder}, we design a Hawkes process-based relation graph convolutional network (HRGCN) as $En()$ to pass message and update entity embeddings on the TKG.

As shown in Fig. \ref{FigHRGCN}, the improvement of HRGCN over traditional RGCN is that HRGCN can effectively learn the structural information from neighbors and assign weights of the messages which represents the temporal decaying of these messages, i.e.
\begin{equation}
    h_o^t = \sigma\Bigg( W_{1} h_o\\
    + \sum\limits_{s,r,t'\in\mathcal{F}_o^t}\frac{1}{|\mathcal{F}_o^t|}W_{r} (h_s+h_r) \widetilde{\kappa}(t-t') \Bigg)   \label{HRGCN}
\end{equation}
where $h_s$, $h_o$ and $h_r$ are input embeddings of nodes $s,o$ and edge $r$ got from $\mathrm{H}_C$ and $\mathrm{R}$, respectively; $\mathcal{F}_o^t$ represents historical neighbors of node $o$, which contains all quadruples = $(s,r,o,t')$ where $t'<t$; $W_r$ is the parameter of message passing between $s$ and $o$; $W_1$ is the parameter of self updating of node $o$; $\widetilde{\kappa}(t-t') = \frac{\kappa(t-t')}{\sum_{s,r,o,t''\in \mathcal{F}_o^t}\kappa(t-t'')}$ is the temporal decaying function, where $\kappa(t-t') = \exp(-\delta(t-t'))$. 

This module outputs the updated embedding matrix $\mathrm{H}_T^t$ that contains historical information. 

\subsubsection{Gating Integration}
The learned embeddings by HRGCN get the historical information of entities, while some useful original information directly from the input TKG may be overwritten.  We use a control gate to balance the contribution of the two kinds of information. 

Specifically, we employ a fully connected layer to generate a graph embedding $h_g$ according to the embeddings $\mathrm{H}_T^t$ from HRGCN. Then we use another fully connected layer to calculate the weight of $\mathrm{H}_T^t$ according to graph embedding $h_g$. The final embedding matrix is a weighted sum of $\mathrm{H}_T^t$ and the initialized embedding matrix $\mathrm{H}_C$.
\begin{equation}
h_g = \sigma(W_{graph}\mathrm{H}_T^t + b_{graph})  \label{graph_embed}
\end{equation}
\begin{equation}
\gamma = \sigma(W_{gate} h_g + b_{gate})  \label{gate}
\end{equation}
\begin{equation}
\mathrm{H}^t = \gamma * \mathrm{H}_T^t + (1-\gamma) \mathrm{H}_{c} \label{weight_sum}
\end{equation}
where $h_g$ is the generated graph embedding, $W_{graph}, W_{gate}, b_{graph} $ and $ b_{gate}$ are learnable parameters.

The output of the encoding module is the balanced evolutional embedding matrix $\mathrm{H}^t$ that encodes the evolutional information on graph $G_{[t_1,t]}$.

\subsection{Conditional Decoding}
The last step of our model is to do reasoning according to the embedding matrix and get the conditional intensity scores of the candidate entities, then select the entity with the highest score as the result, as described in Eq.\ref{decoder}.
\subsubsection{Previous RGCN Decoding}
Existing works \citep{REGCN, HiSMatch} use ConvTransE as the decoding function $De()$ for traditional RGCN to calculate the certain intensity values $\lambda_o$ of entity $o$ by Eq. \ref{convtranse_original}.
\begin{equation}
\lambda_o^{ori} = \mathrm{ConvTransE}(concat(h_{s_q}^{t_q}, h_{r_q}); h_{o_q}^{t_q})  \label{convtranse_original}
\end{equation}

Note that the parameters of ConvTransE are fixed over different queries, which leads the model to tend to reason using the few evolutional patterns of the most common events, causing bias against other evolutional patterns of the non-common events. We adjust ConvTransE to avoid the bias caused by  long-tailed distribution in the following.

\subsubsection{Avoiding Long-tailed Distribution Bias}
We use feature linear modulation (FiLM) to construct a hyper-network, which adjusts parameters of decoder according to different queries, so that it can choose the appropriate query-specific evolutional pattern for reasoning. 

Specifically, given a query quadruple = $(s_q,r_q,o_q,t_q)$, the hyper-network generates a shifting factor $\alpha^{(s_q,r_q,t_q)}$ and a scaling factor $\beta^{(s_q,r_q,t_q)}$ according to the vector of the query quadruple to scale and shift the decoding parameters.
\begin{equation}
\alpha^{(s_q,r_q,t_q)}=\sigma((h_{s_q}^{t_q}||h_{r_q})W_\alpha + b_\alpha)  \label{scale}
\end{equation}
\begin{equation}
\beta^{(s_q,r_q,t_q)}=\sigma((h_{s_q}^{t_q}||h_{r_q})W_\beta + b_\beta)  \label{shift}
\end{equation}
\begin{equation}
\theta^{q}=(\alpha^{(s_q,r_q,t_q)}+1)\odot\theta+\beta^{(s_q,r_q,t_q)}  \label{adjust}  
\end{equation}
where $h_s^{t_q}$ and $h_{r_q}$ are embeddings of subject $s_q$ at time $t_q$ and relation $r_q$ respectively; $W_\alpha, W_\beta, b_\alpha$ and $b_\beta$ are learnable parameters; $\theta$ is the original parameters from decoder and $\theta^{q}$ is query-specific parameters; $\odot$ is Harmard product.

\subsubsection{Adjusted Decoding}
The adjusted decoder extracts the multi-dimensional features of the query quadruple through one-dimensional convolution and gets the conditional intensity scores of candidate entities via the inner product with the embedding matrix.
\begin{equation}
\lambda_o = \mathrm{ConvTransE}([h_{s_q}^{t_q}, h_{r_q}]; h_{o_q}^{t_q};\theta^{q})  \label{convtranse}
\end{equation}
$\lambda_o$ is the conditional intensity of candidate entity $o$; [$\cdot$] means the concat function; $\theta^{q}$ is the adjusted parameters got from the feature transform unit. The decoder chooses the entity which has the highest score as the missing part of the query quadruple.

\subsection{Learning Objective}
The task to predict the missing entity of a given quadruple could be seen as a multi-classification task and each entity in the candidate set belongs to one class. The optimization objective of entity prediction task is to maximum the scores of the ground truth entities, which can be convert to a cross-entropy loss $\mathcal{L}_e$:
\begin{equation}
\mathcal{L}_e = -\sum_{t_q\in \mathcal{T}}{\sum_{\mathcal{F}_{t_q}}{\sum_{k=1}^K y_k \log{p(o_k|s_q,r_q,t_q)}}}    \label{loss_e}
\end{equation}
where $\mathcal{T}$ is the timestamp set; $\mathcal{F}_{t_q}$ is the quadruple set with timestamp $t_q$; $K$ is the number of entities; $y_k=1$ if entity $o_k$ equals to ground truth $o_q$, otherwise 0; $p(o_k|s_q,r_q,t_q)$ is the probability of $o_k$, normalized by the scores of all the candidate entities in the candidate sets.



\section{Experiments}

\subsection{Experimental Setup}

\subsubsection{Datasets}

We use four benchmark datasets which are generally used in TKG reasoning task to evaluate the effectiveness of HERLN, ICEWS14 \citep{REGCN}, ICEWS18 \citep{REGCN}, WIKI \citep{RENET} and YAGO \citep{RENET}. ICEWS is a database got from more than 100 data sources over more than 250 countries and regions. ICEWS14 and ICEWS18 datasets contain events occurred in 2014 and 2018 respectively. WIKI and YAGO are subsets of the Wikipedia history and YAGO3 respectively. We list the statistics of these datasets in Appendix \ref{sec:Statistics of the Datasets}.

\subsubsection{Baselines}

We compare HERLN with 10 TKG reasoning models, which can be categorized into three classes. 
(1) Embedding-based models, RE-NET \citep{RENET}, REGCN \citep{REGCN}, EvoKG \citep{EvoKG}, CENET \citep{CENET}, HGAT \citep{HGAT} and TiPNN \citep{TiPNN}; 
(2) Path-based models, TG-Tucker \citep{NeuralODE} and TLogic \citep{TLogic}; 
(3) Hawkes process-based models, GHNN \citep{GHNN} and GHT \citep{GHT}.

\subsubsection{Evaluation Metrics}

We report MRR, which is the mean of the reciprocal values of the actual missing entities’ ranks averaged by all the queries, and Hits@1/3/10, i.e., the proportion of correct test cases that are ranked within top 1/3/10.

\subsubsection{Implementation Details}

We implement our model in Pytorch \citep{pytorch} and DGL Library \citep{DGL}. The experiments are conducted on a Nvidia GeForce Titan GPU. To be consistent with the baselines, we set the embedding dimension of entities $d_e$ and relations $d_r$ to 200. The number of HRGCN layers is set to 2 and the dropout rate for each layer is set to 0.2. We set all weights of edges to 1 in the embedding initializing module. We use the same hyperparameter settings of ConvTransE given by \citet{REGCN}, the decode unit has 50 convolutional kernels with a size of 2$\times$3 for each kernel. Adam \citep{Adam} is adopted for parameter learning with the learning rate of 0.001 on all the datasets. We report the average experimental results on three random seeds.
Our code is available at https://github.com/WisdomMLlab/HERLN.



\subsection{Experimental Results}
\subsubsection{Entity Prediction}

\begin{table*}[tbh]
\begin{center}
\resizebox{2.0\columnwidth}{!}{
	\begin{tabular}{lcccccccccccccccc}
		\hline
		\multicolumn{1}{c}{\multirow{2}{*}{Method}} & \multicolumn{4}{c}{ICEWS18}       & \multicolumn{4}{c}{ICEWS14} & \multicolumn{4}{c}{WIKI} & \multicolumn{4}{c}{YAGO}      \\ \cline{2-17} 
		\multicolumn{1}{c}{}                        & MRR   & Hits@1 & Hits@3 & Hits@10 & MRR   & Hits@1 & Hits@3 & Hits@10   & MRR   & Hits@1 & Hits@3 & Hits@10 & MRR   & Hits@1 & Hits@3 & Hits@10 \\ \hline
		GHNN                                        & 27.93 & 18.77  & 31.55  & 45.80   & 37.54 & 26.65  & 41.83  & 52.66  & 59.69 & 57.25 & 60.93  & 63.99   & 63.17 & 58.41 & 65.45  & 72.18   \\
        GHT                                         & 27.40 & 18.08  & 30.76  & 45.76   & 37.40 & 27.77  & 41.66  & 56.19    & 60.02 & 59.43 & 61.52 & 63.16 & 68.26 & 60.78 & 70.37 & 79.93 \\
		RE-NET                                      & 28.81 & 19.05  & 32.44  & 47.51   & 38.28 & 28.68  & 41.34  & 54.52   & 49.66 & 46.88 & 51.19  & 53.48   & 58.02 & 53.06 & 61.08  & 66.29   \\
		RE-GCN                                      & 30.55 & 20.00  & 34.73  & \underline{51.46}   & 41.50 & 30.86  & 46.60  & \underline{62.47}    & 51.53 & - & 58.29  & 69.53   & 63.07 & -  & 71.17  & 82.07   \\
		TG-Tucker                                   & 28.68 & 19.35  & 32.17  & 47.04   & 26.25 & 17.30  & 29.07  & 44.18   & 50.43 & 48.52 & 51.47  & 53.58   & 57.83 & 53.05 & 60.78  & 65.85   \\
        TLogic                                       & 29.82 & 20.54  & 33.95  & 48.53   & 43.04 & 33.56  & 48.27  & 61.23    & - & - & -  & -   & - & - & -  & -   \\
		EvoKG                                       & 27.20 & 17.61 & 31.14 & 45.82 & 35.78 & 26.82 & 39.75 & 52.90 & 67.68 & 55.02 & \underline{79.48} & \underline{84.03} & 68.87 & 54.47 & 81.22 & \underline{89.81} \\ 
        CENET                                       & 27.75 & 18.92  & 32.08  & 46.19   & 38.24 & 28.82  & 42.14  & 56.82    & 62.67 & 59.18 & 65.31  & 67.90   & 65.15 & 56.96 & 68.13  & 70.35   \\ 
        HGAT                                       & 28.50 & 19.60  & 32.70  & 46.60   & 38.90 & 29.70  & 42.40  & 56.40    & 56.10 & 52.90 & 58.10  & 61.80   & 63.60 & 59.80 & 66.00  & 71.50   \\ 
        TiPNN                                       & 30.32 & 21.55  & 35.06  & 50.08   & 41.20 & 32.75  & 46.23  & 59.60    & 73.99 & \underline{71.57} & 76.82  & 80.67   & \underline{80.30} & \underline{78.85} & \underline{82.10}  & 89.04   \\ \hline
        OurModel                                    & \textbf{31.33} & \textbf{21.93}  & \textbf{35.59}  & \textbf{52.01}   & \textbf{43.94} & \textbf{34.62}  & \textbf{49.48}  & \textbf{63.44}    & \textbf{79.10}  & \textbf{74.92} & \textbf{82.47} & \textbf{85.31}   & \textbf{84.47} & \textbf{80.31} & \textbf{87.56}  & \textbf{91.06}   \\ \hline
	\end{tabular}
}
\caption{Entity prediction results. The best results are marked in bold and the second best ones are underlined.}
\label{results1}
\end{center}
\end{table*}

Table \ref{results1} shows the entity prediction results on the benchmark datasets. The best results are marked in bold and the second best ones are underlined. 
It can be seen that our proposed HERLN performs the best nearly on all the settings, an only exception is that it achieves the second best Hits@3 score on ICEWS14. 
GHNN and GHT combines Hawkes point process with neural networks. But they do not consider the temporal decaying of event impact, which limits their performance. RE-GCN uses RNN on static graphs and loses temporal information during long-term evolution. CENET measures the probability of different entities by constructing the historical vocabulary. It does not exploit the graph structural information. TiPNN focuses on query-aware temporal path to capture the feature and doesn't take the potential relations between entities into account.

Our model can capture the historical structure and the event evolutional patterns at the same time, considering the community structure, long-tailed distribution, and temporal decaying characteristics, thus it makes significant improvements over the baselines on all the datasets. 

And among these four datasets, our method exhibits significant improvements compared to the baseline on the WIKI and YAGO dataset, while the enhancement on the ICEWS dataset is not as pronounced. This discrepancy is due to the fact that the WIKI and YAGO data both have a temporal span of one year, whereas ICEWS has a span of 24 hours. This aligns with real-world phenomena because the impact of events diminishes more noticeably over a longer time span, and the decay of short-term events is limited. Consequently, HERLN demonstrates substantial improvement in TKGs with longer time spans and distinct community structures.

\subsubsection{Relation Prediction}
The results of the relation prediction task in terms of the MRR metric are shown in Table \ref{results3}. In the relation prediction task experiment, we do not include TG and RE-NET as baseline because they conduct only entity prediction. It can be seen from  Table \ref{results3} that our proposed HERLN outperforms all the baselines and receives a boost of up to 10\% in the MRR metric. The reasons of the performances of both our model and the baselines are similar to those for the entity prediction task. 

\begin{table}[tbh]
\begin{center}
\resizebox{1.0\columnwidth}{!}{
    \begin{tabular}{lcccccc}
    \hline
        \multicolumn{1}{c}{Method}		& ICEWS18 			& ICEWS14s  			& WIKI 				& YAGO				 \\ \hline
        ConvE 		& 37.73 			& 38.80 			& 78.23 			& 91.33 			\\
        ConvTransE	& 38.00 			& 38.40 			& 86.64 			& 90.98 			\\
        R-GCRN		& 37.14 			& 38.04 			& 88.88 			& 90.18 				\\
        RE-GCN 		& 40.53         	& 41.06			    & \underline{97.92} & \underline{97.74}				\\
        EvoKG 		& \underline{41.11} & \underline{42.47}	& 90.63 			& 90.26				\\
        CENET 		& 38.24 	        & 40.50			    & 87.51 			& 91.44				\\
        OurModel 	& \textbf{51.47} 	& \textbf{50.55} 	& \textbf{98.50} 	& \textbf{98.54} 	\\ \hline
    \end{tabular}
}
\caption{Relation prediction results by MRR metric.}
\label{results3}
\end{center}
\end{table}

\subsubsection{Ablation Study}
We conduct ablation experiments on the ICEWS14 dataset. (1) OurModel without (w.o) ConvTransE: we replace the ConvTransE in the decoder unit with a simple MLP layer. (2) OurModel  without (w.o) FiLM: Instead of FiLM, we directly use ConvTransE. (3) OurModel  without (w.o) HRGCN: we replace HRGCN with a RGCN to aggregate snapshots. (4) OurModel without (w.o) Community: we omit the embedding initialize module. 


As shown in Table \ref{ablation}, replacing any component of our model degrades the performance, which demonstrates that each component of the model has a positive gain on the result. 
The variant (1) has an almost 6\% drop in MRR, suggesting that an appropriate decoder can learn event evolution patterns effectively. The variant (2) shows 2\% decreasing in Hits@1, indicating that the hypernetwork helps the model distinguish different event evolution patterns. 
The variant (3) has 7.89\%, 6.89\%, 9.31\%, 11.90\% drops in MRR, Hits@1, Hits@3, respectively, emphasizing the importance of incorporating the temporal information of events through HRGCN. The variant (4) has 3\% decreasing in MRR, confirming the helpfulness of extracting community structures.

\begin{table}[tbh]
\begin{center}
\resizebox{1.0\columnwidth}{!}{
    \begin{tabular}{cccccc}
    \hline
        Method	 			& MRR			 	& Hits@1 	& Hits@3 	& Hits@10 	\\ \hline
        w.o ConvTransE		& 37.98		& 29.59 			& 41.72 			& 54.19 			\\
        w.o FiLM 		& 41.48 		& 31.39 			& 46.47 			& 60.14 			\\
        w.o HRGCN 	& 36.05 		& 27.73 			& 40.17 			& 51.54 			\\
        w.o Community 		& 41.05 		& 31.86 			& 45.96 			& 58.25 			\\
         \textbf{OurModel} 			& \textbf{43.94} 		& \textbf{34.62} 			& \textbf{49.48} 			& \textbf{63.44} 			\\ \hline
    \end{tabular}
}
\caption{Effect of main components}
\label{ablation}
\end{center}
\end{table}
\section{Conclusions}
In this paper, we propose a Hawkes-based evolutional representation learning network (HERLN) to tackle the TKG reasoning tasks. We exploit the structural and evolutional characteristics of TKGs inherited from real-world networks to learn structural information and evolutional patterns. We initialize the embeddings with a community detection algorithm and a graph convolution network to make use of community structure. We design a Hawkes process-based relational graph convolutional network to tackle the temporal decaying phenomenon. We construct a conditional decoder to alleviate the biases caused by scale-free property (long-tailed distribution). Experimental results show the superiority of our proposed model. 

\section{Limitations}
Note that there are many entities appear only in the test set (named unseen entities), which will continue to appear as the size of TKG increases. Our model could not get sufficient information to assign proper embbedings for these entities. For further work, we plan to find a method to deal with the unseen entities.

\section*{Acknowledgments}
This work was supported by National Natural Science Foundation of China (No. 62476040) and the Fundamental Research Funds for the Central Universities.

\bibliography{custom}

\section*{Appendix}
\appendix

\section{Statistics of the Datasets}
\label{sec:Statistics of the Datasets}
The statistics of the datasets used in our experiment are summarized in Table \ref{statistics}. We divide ICEWS14, ICEWS18, WIKI and YAGO into training, validation and test sets with a proportion of 80\%, 10\% and 10\% in the chronological order, i.e., $t_{train}<t_{valid}<t_{test}$. 

\begin{table*}[htb]
\begin{center}
    \begin{tabular}{lcccccccc}
    \hline
        \multicolumn{1}{c}{\textbf{Dataset}} & $\boldsymbol{\mathcal{E}}$ & $\boldsymbol{\mathcal{R}}$ & $\boldsymbol{\mathcal{F}}$ & $\boldsymbol{\mathcal{T}}$ & $\boldsymbol{\mathcal{F}_{train}}$ & $\boldsymbol{\mathcal{F}_{valid}}$ & $\boldsymbol{\mathcal{F}_{test}}$ &  \textbf{Time interval} \\ \hline
        ICEWS18 	& 23033 	& 256 	& 468558 	& 304 & 373019 & 45996 & 49546	& 24 hours \\ 
        ICEWS14 	& 7128 	& 230 	& 90730 	& 365 & 74846 & 8515 & 7372	& 24 hours \\ 
        WIKI 		& 12554 	& 24 		& 669934 	& 231 & 539287 & 67539 & 63111	& 1 year \\ 
        YAGO 		& 10623 	& 10 		& 201089 	& 187 & 161541 & 19524 & 20027 & 1 year \\ \hline
    \end{tabular}
\caption{Statistics of the datasets}
\label{statistics}
\end{center}
\end{table*}

\section{Case Study}
In Table \ref{case study}, we present three cases obtained from the ICEWS14s test set, each of which validates the effectiveness of one contribution within HERLN. In all three cases, HERLN achieves the highest score and successfully completes the predictions. In the first case, two related events share the same subject entity and relation, with a different object entity. HERLN leverages the community structure of Criminal (Venezuela) and Citizen (Venezuela) to complete the reasoning. In the second case, multiple events occurre between three entities, but the influence of earlier events on the reasoning outcome 10 months later is evidently smaller than that of more recent events. The Hawkes process helps our model to distinguish the useful information and make the correct choice. In the last case, \textit{Demand settling of dispute} is a rare relation type, occurring less than 10 times in the dataset. The model is prone to be influenced by more common evolution patterns during reasoning. HERLN still successfully learns this evolution pattern by constructing a hyper network, completing the reasoning.

\begin{table*}[tbh]
\begin{center}
\resizebox{2.0 \columnwidth}{!}{
    \begin{tabular}{lll}
        \hline
            \textbf{Query} & \textbf{Answer} & \textbf{History } \\ \hline
            \makecell[l]{Criminal (Venezuela), fight with small \\ arms and light weapons, ?, 2014/11/20} & Citizen (Venezuela)  & \makecell[l]{- Criminal (Venezuela), Use unconventional \\  violence, Citizen (Venezuela), 2014/3/17 \\- Criminal (Venezuela), Use unconventional \\  violence, Businessperson (Germany), 2014/6/19} \\ 
            Jason C. Hu, Yield, ?, 2014/11/29 & Lin Chia-lung  & \makecell[l]{- Jason C. Hu, Praise or endorse, Lin \\Chia-lung, 2014/11/25 \\- Jason C. Hu, consult, Ma Ying Jeou,2014/1/16} \\ 
            \makecell[l]{John Kerry, Demand settling \\of dispute, ?, 2014/11/12} & Iran  & - John Kerry , Make statement,Iran,2014/11/9 \\ \hline
    \end{tabular}
}
\caption{Case Study}
\label{case study}
\end{center}
\end{table*}

\section{Time Cost Analysis}
\label{Time Cost Analysis}
\begin{figure}[tbhp]
\centering
\includegraphics[width=1.0\columnwidth]{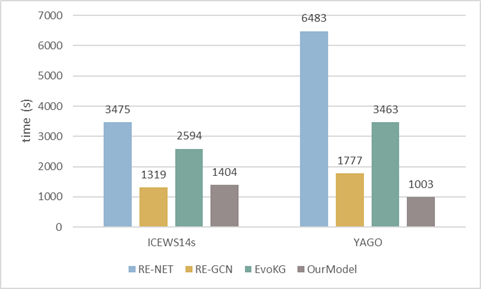}
\caption{Time cost analysis on ICEWS14s and YAGO.}
\label{timecost}
\end{figure}

To evaluate the efficiency of our model, we compare HERLN with several TKG reasoning methods including RE-NET, RE-GCN and EvoKG. We select these baseline methods for the time analysis because these methods are similar to ours, being embedding-based methods. Comparing the time efficiency on these methods can better illustrate the efficiency improvement of our approach. CENET is an exception as it does not provide official code. Although we conduct the experiments, the computational time is significantly longer than other methods (exceeding 1 day on ICEWS14s). Therefore, we exclude CENET from the results. As Fig. \ref{timecost} shows, our model runs faster than RE-NET and has a similar time consumption with RE-GCN. On the one hand, RE-NET uses multiple RNN structures to fit multi-level conditional probability distributions of events while ours relies on a Hawkes process for message passing and aggregation on one single graph. On the other hand, we use query-specific version of ConvTransE as the decoder. The structure of ConvTransE allows it to predict multiple events at the same time, this high parallelism saves much time.

\end{document}